\documentclass[10pt,journal]{IEEEtran}

\usepackage[T1]{fontenc}
\usepackage{filecontents}

\usepackage{xr}
\usepackage{graphicx}

\usepackage{cite}

\usepackage{standalone}

\usepackage{pgfplots}
\pgfplotsset{compat=newest}

\usepackage{pgfplotstable}
\usepgfplotslibrary{external} 
\usetikzlibrary{patterns}
\usepackage{circuitikz}

\usepackage{subfigure}
\usepackage{amsmath}
\usepackage{amsfonts}
\usepackage{amssymb}
\usepackage{wasysym}

\usepackage{array}

\usepackage{siunitx}
\renewcommand\Re{\operatorname{Re}}
\renewcommand\Im{\operatorname{Im}}

\newcommand{\FigWidth}{7.5cm}
\newcommand{\FigHeight}{5.5cm}

\usepackage{stfloats}

 \usepackage{booktabs}
 \usepackage{multirow}
 \usepackage{graphicx}

\begin{document}

\graphicspath{{Fig-pdf/}}

\setcounter{page}{4686}

\title{The impact of reduced conductivity on the performance of wire antennas}

\author{Morteza~Shahpari,~\IEEEmembership{Member,~IEEE,} David~V.~Thiel,~\IEEEmembership{Senior~Member,~IEEE,} \thanks{M. Shahpari and D. Thiel are with Griffith School of Engineering, Griffith University, Nathan, Queensland 4111, Australia} \thanks{This work is partly funded by the grant number  DP130102098 from Australian Research Council.}
\thanks{Manuscript contains supplementary material which are available online.}
}

\markboth{IEEE Transactions on Antennas and Propagation,~Vol.~63, No.11, November~2015}{Shahpari, Thiel: The impact of reduced conductivity on the performance of wire antennas}

\maketitle

\begin{abstract}
Low cost methods of antenna production primarily aim to reduce the cost of metalization. This might lead to a reduction in conductivity. 
A systematic study on the impact of conductivity is presented.
The efficiency, gain and bandwidth of cylindrical wire meander line, dipole, and Yagi-Uda antennas were compared for materials with conductivities in the range $\mathrm{10^3}$ to $\mathrm{10^9}$ S/m. 
In this range, the absorption efficiency of both the dipole and meander line changed little, however the conductivity significantly impacts on radiation efficiency and the absorption cross section of the antennas. 
The extinction cross section of the dipole and meander line antennas (antennas that Thevenin equivalent circuit is applicable) also vary with radiation efficiency. 
From the point of radiation efficiency, the dipole antenna performance is most robust under decreasing conductivity.
Antennas studied in this study were fabricated with brass and graphite.
Radiation efficiency of the antennas were measured by improved Wheeler cap (IWC) method.
Measurement results showed a reasonable agreement with simulations.
We also measured the extinction cross section of the six fabricated prototypes.

\end{abstract}


\section{Introduction}

\IEEEPARstart{M}{aterials} like carbon nanotubes (CNT) \cite{DeVolder_2013_Sci,Behabtu_2013_Sci} , and printed conductor technologies like circuits in plastic (CiP) \cite{Thiel_2009_EPTC,Thiel_2008_ISAPE} are finding their way to telecommunications systems. 
These applications are in demand because of the low cost of production and ability to produce light weight, recyclable and flexible circuits. 
However, the conductivity of printed conductors is usually much lower than the conductivity of copper, aluminum and silver. 
The main aim of this paper is to study the performance of lossy transmitting and receiving antennas made from materials with lower conductivities.

Some RFID designs with reduced conductivity are reported in \cite{Nikitin_2005_APS,Yilmaz_2008_APS}. 
The effect of the conductor thickness on the radiation efficiency and backscatter power of RFIDs are studied in \cite{Siden_2007_MAP,Merilampi_2010_ProcIEEE}. 
Also, a cost study of the printed antennas is available from \cite{Pongpaibool_2012_ISAP} for dipole antenna with different conductivities.
Despite the interesting results presented in \cite{Nikitin_2005_APS,Yilmaz_2008_APS,Siden_2007_MAP,Merilampi_2010_ProcIEEE,Pongpaibool_2012_ISAP}, a comprehensive study of the effects of conductivity on the various parameters of the antennas is missing.
In this paper, we address this gap in the current literature by exploring the effects of the change in the conductivity of different antennas.

There are some advantages in choosing wire rather planar structures for these investigations.
First, wire antennas have been studied extensively in the literature.
Also, new materials like carbon nanotubes are emerging which show promising characteristics and naturally have circular cross sections.
We performed simulations over $10^3-10^9S/m$ to cover materials like CNTs which have conductivities in the order of $10^4-10^7S/m$ \cite{Behabtu_2013_Sci}.

The key point in the success of any mass production process is low cost, that is, circuit manufacturing techniques tend to minimize the total used materials.
By tapering the wire in \cite{Galehdar_2009_AWPL}, the same performance was achieved with a conductor volume reduction of more than 50\%.

In this paper, we illustrate the influence of the non perfect materials on the impedance and radiation properties of three wire antennas: a half-wave dipole, a four element Yagi-Uda, and spiral meander line antennas.
Such investigations have not been reported previously.
We also study that how absorbed, scattered, and dissipated power in a receiving antenna changes with conductivity in section~\ref{Sec_Num_Results}.
The study reveals the relationship between the absorption cross section of the antennas and the radiation efficiency. 
We demonstrate that the extinction cross section of the dipole and meander line antennas is directly related to the radiation efficiency of these antennas.
Moreover, we show the impact of conductivity on absorption efficiency, generalized absorption efficiency, and absorbed power to volume ratio in section~\ref{Sec_Num_Results} with additional data in section~\ref{Sec_SI_Graphs}.
Section~\ref{Sec_Fabrication} shows the fabricated samples.
Measurement results are discussed in section~\ref{Sec_MeasRes}.

\section{Antenna Designs}
\label{Sec_AntDes}
In this paper, we selected three test case antennas: a half-wave dipole, a spiral meander line, and a Yagi-Uda antenna. 
These three wire antennas were designed to resonate at the same frequency ($f=$\SI{1}{\giga\hertz}). 
The dipole antenna was selected because directivity and absorption efficiency are available in the closed form. 
Meander line antennas are one of the most popular choices for small antenna designs. 
The Yagi-Uda antenna was included in this study because of their exceptional absorption efficiency. 
The geometry of the meander line and Yagi-Uda antenna are selected from the  \cite{Shahpari_2014_TAP,Goudos_2010_PIER}, respectively. 
The Yagi-Uda antenna was optimized in terms of gain, side lobe level (SLL), and voltage standing wave ratio (VSWR).

\begin{figure}
\centering
\subfigure [ ]{
\begin{tikzpicture}
\draw [thick] (-0.1,0.1) -- (0,0.1) -- (0,1);
\draw [thick] (-0.1,-0.1) --(0,-0.1) -- (0,-1);
\draw [<->,dashed] (0.25,-1)--(0.25,1);
\node [rotate=-90] at (0.5,0) {143};
\end{tikzpicture}
}
\subfigure[ ]{
\begin{tikzpicture} [scale=0.054]
\draw(9.25,12.5) -- (9.25,18.75) -- (15.5,18.75) -- (21.75,18.75) -- (21.75,12.5) -- (21.75,6.25) -- (15.5,6.25) -- (9.25,6.25) -- (3,6.25) -- (3,12.5) -- (3,18.75) -- (3,25) -- (9.25,25) -- (15.5,25) -- (21.75,25) -- (28,25) -- (28,18.75) -- (28,12.5) -- (28,6.25) -- (28,0) -- (21.75,0) -- (15.5,0) -- (9.25,0) -- (3,0)
 -- (1,0) -- (1,-2) (-1,-2) -- (-1,0) -- 
 (-3,0) -- (-9.25,0) -- (-15.5,0) -- (-21.75,0) -- (-28,0) -- (-28,6.25) -- (-28,12.5) -- (-28,18.75) -- (-28,25) -- (-21.75,25) -- (-15.5,25) -- (-9.25,25) -- (-3,25) -- (-3,18.75) -- (-3,12.5) -- (-3,6.25) -- (-9.25,6.25) -- (-15.5,6.25) -- (-21.75,6.25) -- (-21.75,12.5) -- (-21.75,18.75) -- (-15.5,18.75) -- (-9.25,18.75) -- (-9.25,12.5);
\draw [<->,dashed] (-28,27)--(28,27);
\node at (0,31) {32.31};
\draw [<->,dashed] (30,0)--(30,25);
\node [rotate = -90]at (35,14) {14.42};
\end{tikzpicture}
}
\subfigure [ ] {
\begin{tikzpicture}[scale =3.5]
\draw (0,-0.236) -- (0,0.236);
\draw (0.294,-0.232) -- (0.294,-0.025);
\draw (0.294,0.232) -- (0.294,0.025);

\draw (0.27,-0.025) -- (0.294,-0.025);
\draw (0.27,0.025) -- (0.294,0.025);

\draw (0.514,-0.220) -- (0.514,0.220);
\draw (0.782,-0.211) -- (0.782,0.211);

% Dimensions Length of the wires
\draw [dashed, <->] (0.07,-0.236) -- (0.07,0.236);
\node [rotate = -90] at (0.12,0) {141.5};
\draw [dashed, <->] (0.34,-0.232) -- (0.34,0.232);
\node [rotate = -90] at (0.40,0) {139.1};
\draw [dashed, <->] (0.56,-0.22) -- (0.56,0.22);
\node [rotate = -90] at (0.62,0) {131.9};
\draw [dashed, <->] (0.83,-0.211) -- (0.83,0.211);
\node [rotate = -90] at (0.89,0) {126.5};

\draw [blue, <->](0,-0.3) -- (0.294, -0.3) node [midway, fill = white, rotate = -90] {88.13};

\draw [blue, <->](0.294,-0.3) -- (0.514, -0.3) node [midway, fill = white, rotate = -90] {65.95};

\draw [blue, <->](0.782,-0.3) -- (0.514, -0.3) node [midway, fill = white, rotate = -90] {80.34};

%s2 88.13   s3  65.95  s4 80.34

\end{tikzpicture}
}
\caption{ Self resonant antenna design configurations at $f_0 = $\SI{1}{\giga\hertz} (a) a half wave dipole (b) a meander line antenna (c) a Yagi-Uda antenna with one reflector and two directors
(all dimensions are in mm).} 
\label{Fig_AntennDesign}
\end{figure}
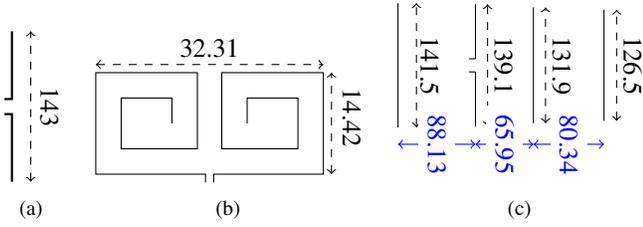

To make these three antennas comparable with each other, we selected a wire radius of $0.00225\lambda$ for all antennas. 
The three antenna designs are shown in Fig.~\ref{Fig_AntennDesign}.

\section{Numerical Results}
\label{Sec_Num_Results}
The antennas were simulated in free space using a commercial method of moments (MoM) code~\cite{FEKO} by changing the conductivity over the range $10^3-10^9$S/m.
It should be noted that the geometric parameters for each antenna over the simulations were similar, while the material properties (conductivity of the metal) are changed.

\begin{figure}
\centering
\subfigure[PEC]{
\includegraphics[width=0.5\linewidth]{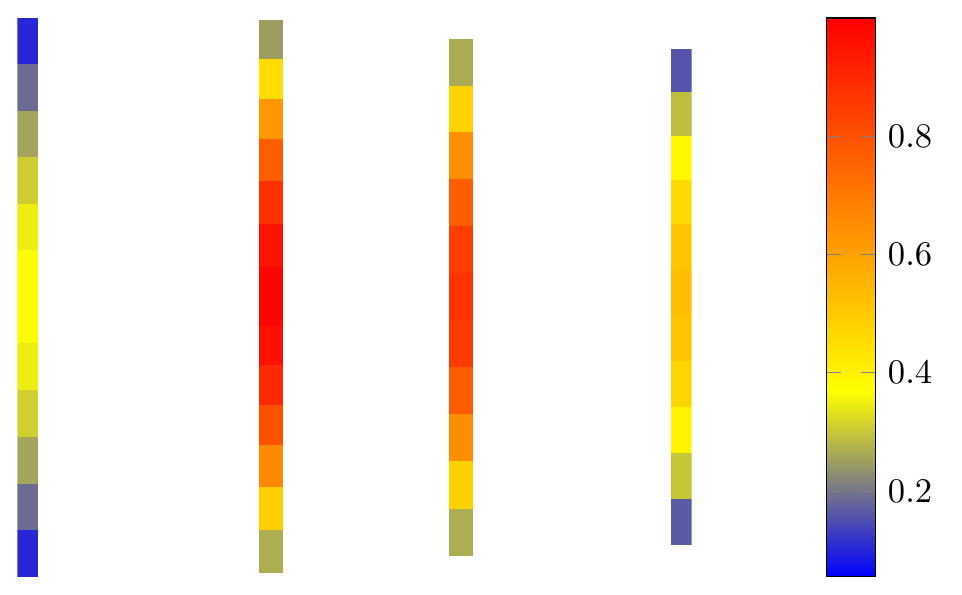}
}
\subfigure[$\sigma=10^3$S/m]{
\includegraphics[width=0.37\linewidth]{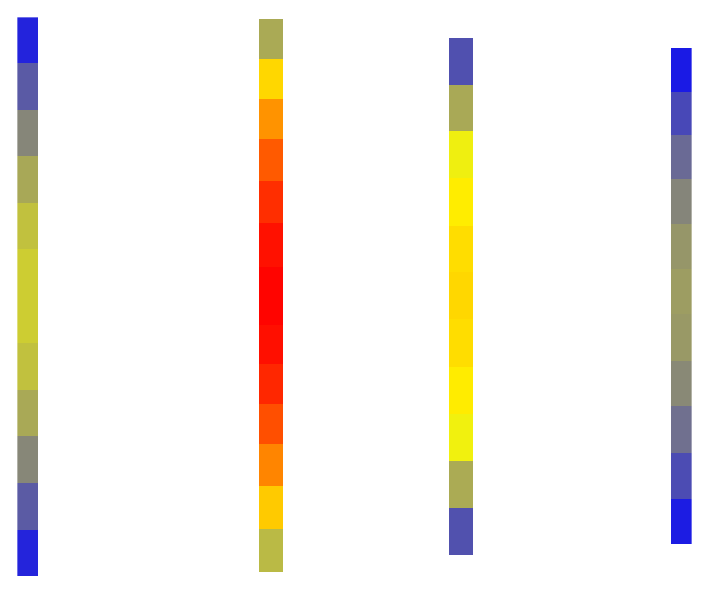}
}
\caption{Comparison of current distribution over  the Yagi-Uda antenna made of (a) lossless (b) lossy ($\sigma=10^3$S/m) materials}
\label{Fig_CurrentYagi}
\end{figure}

Fig.~\ref{Fig_CurrentYagi} shows a comparison of the current distribution of lossy and lossless Yagi-Uda.
In order to make a meaningful comparison, quantities are normalized to the values at the feed point of the antenna.
Fig.~\ref{Fig_CurrentYagi} shows that the normalized magnitude  of induced currents on the elements of Yagi-Uda are different for lossless and lossy cases.
As an example, the normalized magnitude of the current distribution is almost the same for the driven element radiators but the induced current is significantly less (almost half) on the parasitic elements when the material is changed from PEC down to conductivity $10^3$S/m.
Current distribution of the dipole and meander line was not affected significantly with the reduction of the conductivity to $10^3$S/m. 
For instance, lossy dipole still shows a sinusoidal current distribution though with a much lower magnitude.

\begin{figure}
\centering
\includegraphics[width=0.95\columnwidth]{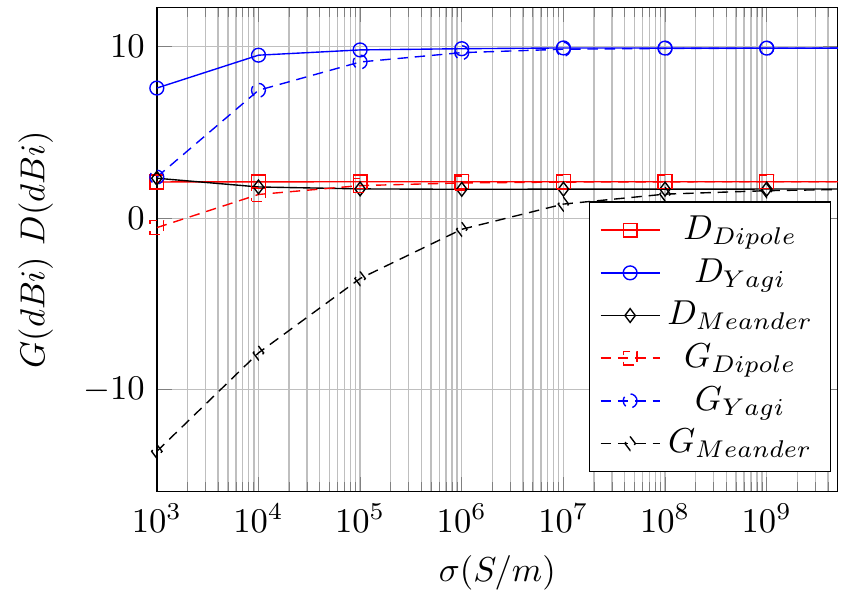}
\caption{Directivity $D$ (solid) and gain $G$ (dashed) versus conductivity $\sigma$ (subscripts $D$, $Y$, and $M$ refer to dipole, Yagi-Uda, and meander line antennas respectively).}
\label{Fig_D_sigma}
\end{figure}

\begin{figure}
\centering

\includegraphics{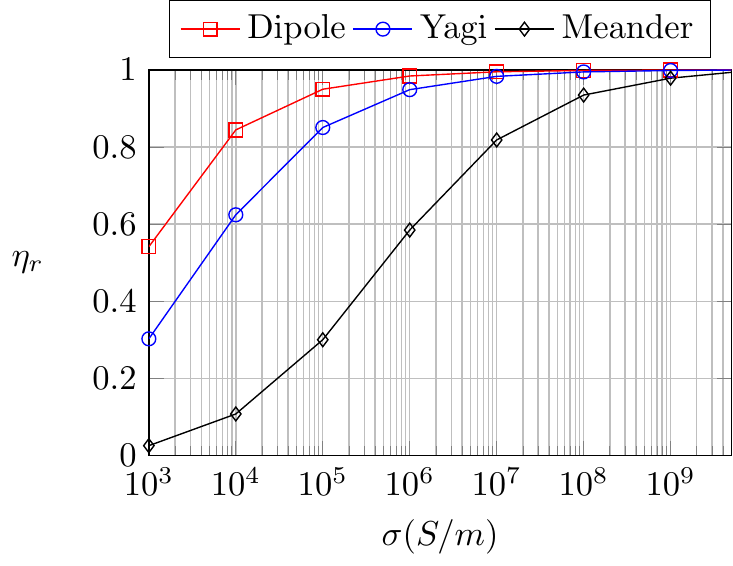}
\caption{Radiation efficiency $\eta_r$ versus conductivity $\sigma$}
\label{Fig_eta_r_sigma}
\end{figure}

% % % % % % % % % % % % % % % % % % % % % % % % % % % % %\
% % % % % % % % % % % % % % % % % % % % % % % % % % % % %
\begin{figure*}
%\centering
\includegraphics[width=\linewidth]{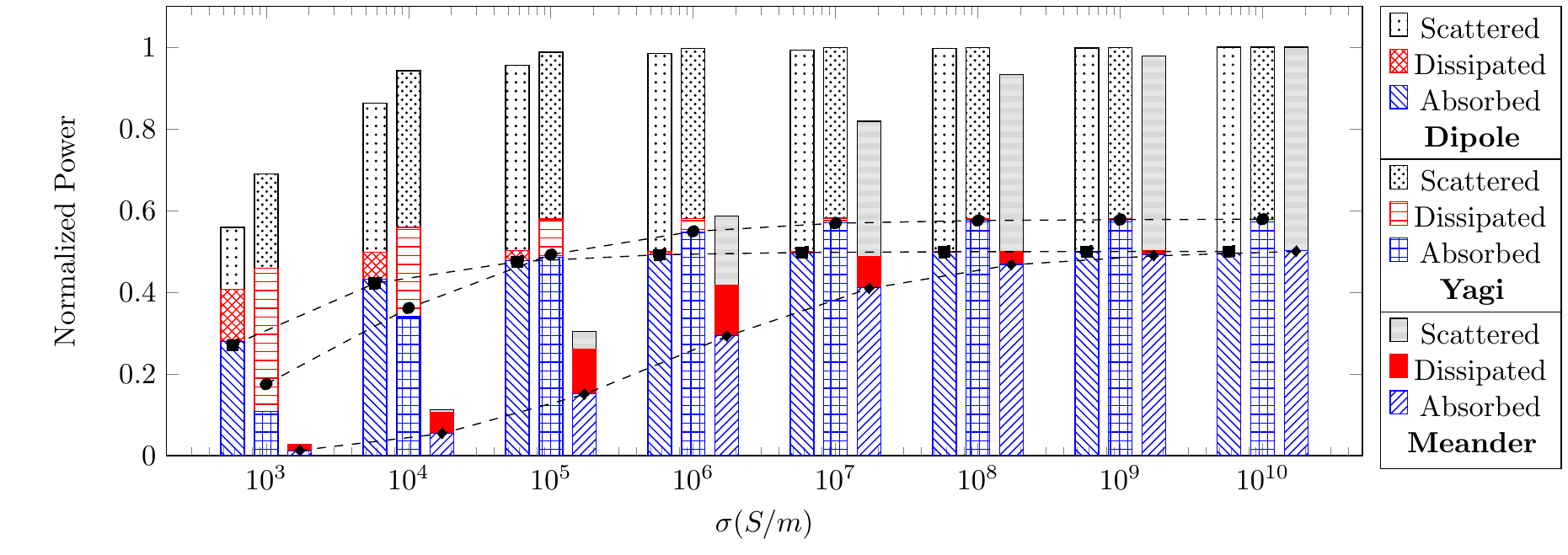}
\caption{Impact of finite conductivity on the absorbed, scattered, dissipated, and extinct powers in antennas. 
Since $P_{ext}=P_a + P_s +P_{loss}$, the height of each bar shows extinct power.
All powers are normalized to the extinct power of PEC antenna of the same type. 
Dashed traces are the product of radiation efficiency and absorbed power by a PEC antenna.}
\label{Fig_Ext_Power}
\end{figure*}

The directivity and gain of the antennas are depicted in Fig.~\ref{Fig_D_sigma}. 
The directivity of the dipole is 2.12dBi which is close to 2.15dBi, the analytical result. 
The directivity $D$ of the dipole remains unchanged while changes are observed for the Yagi-Uda antenna for $\sigma<10^5$(S/m).
It is interesting to see the gain of the dipole does not change much over this conductivity range, however, gain of the meander line falls to negative values for $\sigma\ll 10^7$S/m.

As $G=\eta_r D$, it is expected that $G$ and $\eta_r$ should decrease with the conductivity.
\footnote{Relation $G=\eta_r D$ is separately investigated in Fig.~S2.} 
The value of $\eta_r$ shows that the meander line antenna is more sensitive to less conductive materials (see Fig.~\ref{Fig_eta_r_sigma}).
This is partly explained by the small radiation resistance of the meander line (about $2\Omega$ for PEC).  
For example, ohmic resistance of the meander line with $\sigma \approx 10^7$ S/m is about  $0.5\Omega$, which reduces $\eta_r$ to $0.8$. 
The dipole and Yagi-Uda antennas are less sensitive because of their higher radiation resistance which is $70,50\Omega$, respectively.
The radiation efficiency $\eta_r$ drops for all conductivities when the radiating structure is changed from a dipole to a Yagi-Uda antenna.
This is because parasitic elements in a lossy Yagi-Uda dissipate electromagnetic power while  directing the radiation.

Some of the key results of the paper are illustrated in  Fig.~\ref{Fig_Ext_Power} which shows the performance of lossy antennas in the receiving mode.
Absorbed, dissipated, and scattered powers are stacked on top of each other and normalized to the maximum extinct power for each antenna type made of PEC. 
More power is absorbed and scattered if the radiating structure is made from materials with higher conductivities, and the dissipated power vanishes.
The absorbed power by a lossy antenna is directly influenced by radiation efficiency $\eta_r$. 
This is illustrated by the dashed traces ($\eta_r P_{a,PEC}$) in Fig.~\ref{Fig_Ext_Power} where $P_{a,PEC}$ is the absorbed power by the same antenna made of PEC.
Of course, one can convert absorbed, scattered, extinct power terms to appropriate cross sections.
The increase in extinct power (extinction cross section $\sigma_{ext}$) with conductivity for the antennas is shown in Fig.~\ref{Fig_Ext_Power}.
Graphs to directly compare the cross sections with the product of $\eta_r$ are included in  Fig.~S3- Fig.~S5.
Although a similar trend like $\eta_r$ is observed for $\sigma_a$ and $\sigma_{ext}$, but $\sigma_s$ vanishes more quickly to accommodate for lost power as conductivity decreases.

\begin{figure}
\centering
\includegraphics{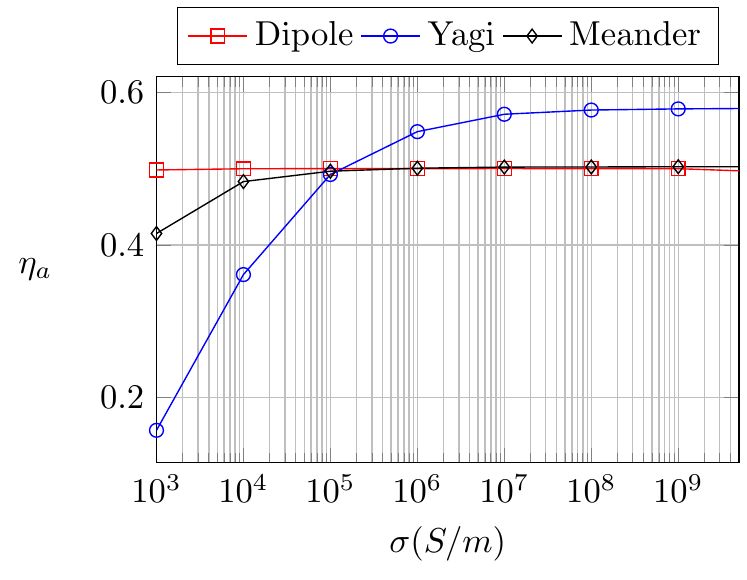}
\caption{Absorption efficiency $\eta_a$ of the antennas versus conductivity $\sigma$}
\label{Fig_eta_a_sigma}
\end{figure}

Absorption efficiency $\eta_a$ is the same for the lossless and lossy dipole antennas (see Fig.~\ref{Fig_eta_a_sigma}). 
This phenomena can be explained in terms of the Thevenin equivalent circuit of the receiving antennas which is discussed in many publications \cite{Collin_2003_APM,Collin_2003_APM_Reply,Andersen_2003_APM,Pozar_2004_APM,Geyi_2004_TAP,Best_2009_APM,Steyskal_2010_APM,Alu_2010_TAP,Liberal_2013_TAP}.
Considering the limitations of the equivalent circuit antenna model \cite{Collin_2003_APM,Best_2009_APM}, we can only apply the Thevenin circuit model to the dipole and meander line antennas. 
We assume that $R_{rad}$, $R_{loss}$, $X_a$, and $Z_L$ are  radiation resistance, loss resistance, antenna reactance, and load impedance respectively. 
The absorption of the maximum power occurs when $Z_L=R_{rad}+R_{loss}-\mathrm{j} X_a$. 
As the conductivity decreases ($R_{loss}$ increases) in a resonant antenna, $Z_L$ is selected appropriately as $R_{rad}+R_{loss}$. 
Therefore, independent of the conductivity, half of the power is absorbed by the antenna load.
Although the absorbed power is reduced by a factor of radiation efficiency $\eta_r$.

Despite observing a similar trend for the Yagi-Uda antenna, it should be noted that the above argument is not appropriate for Yagi-Uda antennas \cite{Andersen_2003_APM,Best_2009_APM}. 
In the other words, one cannot separate scattered and absorbed powers in a simple equivalent circuit, instead one should deal with a multi-port network \cite{Andersen_2003_APM,Liberal_2013_TAP} like Fig S.1.

\begin{figure}
\centering
\includegraphics{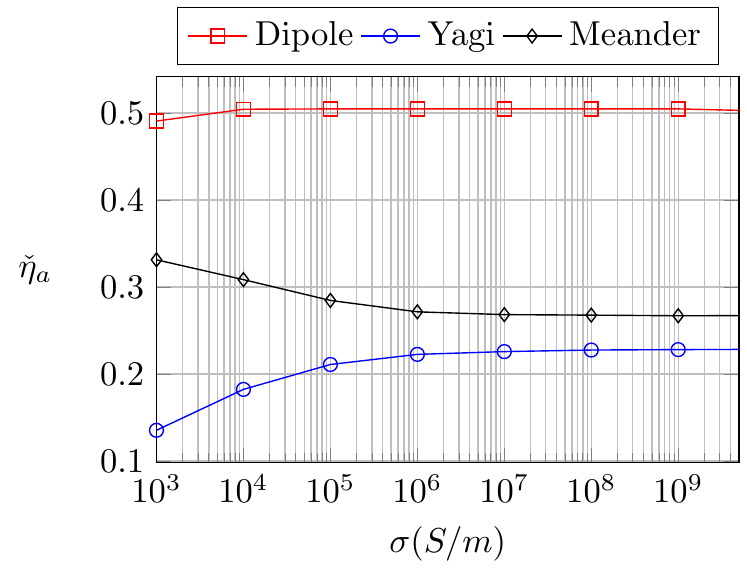}
\caption{Generalized absorption efficiency $\check{\eta}_a$ of the antennas versus conductivity $\sigma$}
\label{Fig_eta_aG_sigma}
\end{figure}

Generalized absorption efficiency $\check{\eta}_a$ was introduced in \cite{Gustafsson_2007_RSPA,Gustafsson_2009_TAP} as the ratio of the total absorbed power to the total extinct power when illuminated by a wideband wave.
This parameter was studied for the meander and Yagi antennas in \cite{Shahpari_2014_TAP,Shahpari_2014_APS}.
Calculation of $\check{\eta}_a$ requires the numerical modeling of the receiving antenna across a broad frequency range. 
We used the frequency range \SIrange{0.1}{20}{\giga\hertz} to ensure the accuracy of results.
It is interesting to note that the antennas have different trends with decreasing conductivity (see Fig.~\ref{Fig_eta_aG_sigma}). 
For instance, the generalized absorption efficiency of the dipole is not sensitive to the change of conductivity, while $\check{\eta}_a$ falls for lossy Yagi-Uda antenna. 
On the other hand, $\check{\eta}_a$ tends to increase for the meander line antenna as the conductivity decreases.

As the conducting material becomes more lossy, $P_{loss}$ increases and the efficiency of the antenna decreases and the bandwidth increases. 
Fig.~\ref{Fig_BW_sigma} shows that for $\sigma<10^5$ S/m the bandwidth increases for all three antennas.
A similar trend is observed for $Q$ factor. 
More specifically, the $Q$ factor of lossy antennas  exactly follows the behavior of the radiation efficiency $\eta_r$. 
That is, $Q$ of the lossy case is equal to the product of $Q$ of lossless case and radiation efficiency $(Q_{lossy}=\eta_r  Q_{lossless})$ (see also  Fig.~S6).
This behavior is exactly what is expected from the theory of fundamental limits on $Q$ factor \cite{Chu_1948_JAP}.

% % % % % % % % % % % % % % % % % % % % % % % % % % % % % % % % % % % % % % % % % % % % % % % % % % % % 
% % % % % % % % % % % % % % % % % % % % % % % % % % % % % % % % % % % % % % % % % % % % % % % % % % % % 
% % % % % % % % % % % % % % % % % % % % % % % % % % % % % % % % % % % % % % % % % % % % % % % % % % % % 

\pgfplotstableread{DataQ.dat} \DataQ
\pgfplotstableread{DataImpBW.dat} \DataImpBW
\begin{figure}
\centering
\begin{tikzpicture}
\pgfplotsset{every axis legend/.append style={
at={(0.5,1.03)},
anchor=south}}
\begin{loglogaxis}[ width=\FigWidth, height =\FigHeight, xlabel=$\sigma (S/m)$, ylabel=$BW (\%)$, %y label style={rotate=-90}, 
xmax=5e9, xmin=1e3,
axis y line = left]

\addplot [red, mark=square] table[x=sigma,y=Dipole] {DataImpBW.dat}; \label{Tr_BW_D};

\addplot [blue, mark=o] table[x=sigma,y=Yagi ] {DataImpBW.dat}; \label{Tr_BW_Y};

\addplot [black, mark=diamond] table[x=sigma,y=Meander] {DataImpBW.dat}; \label{Tr_BW_M};

\end{loglogaxis}
\begin{loglogaxis}[ width=\FigWidth, height =\FigHeight, ylabel=$Q$, 
xmax=5e9, xmin=1e3,
axis y line = right, hide x axis,
y label style = {rotate = -90},
]

\addplot [red, thick,dashed,mark = *]table[x=sigma,y=Dipole] {\DataQ}; \label{Tr_Q_D};

\addplot [blue,dashed, thick,mark=square*] table[x=sigma,y=Yagi] {\DataQ}; \label{Tr_Q_Y};

\addplot [thick,black,dashed, mark = diamond*] table[x=sigma,y=Meander] {\DataQ}; \label{Tr_Q_M};

\end{loglogaxis}

\draw [rotate = 20] (0.85,0.5) ellipse (0.2 and 0.6);
\draw [->, thick] (0.62,1.2) -- (1.5, 1.2);

\draw (0.75,3.) ellipse (0.2 and 0.7);
\draw [->, thick] (0.6,2.5) -- (.15, 2.5);

\end{tikzpicture}

\caption[m]{Bandwidth (dipole \ref{Tr_BW_D}, Yagi \ref{Tr_BW_Y}, and meander line \ref{Tr_BW_M}) and $Q$ factor (dipole \ref{Tr_Q_D}, Yagi \ref{Tr_Q_Y}, and meander line \ref{Tr_Q_M}) of the antennas versus $\sigma(S/m)$. }
\label{Fig_BW_sigma}
\end{figure}
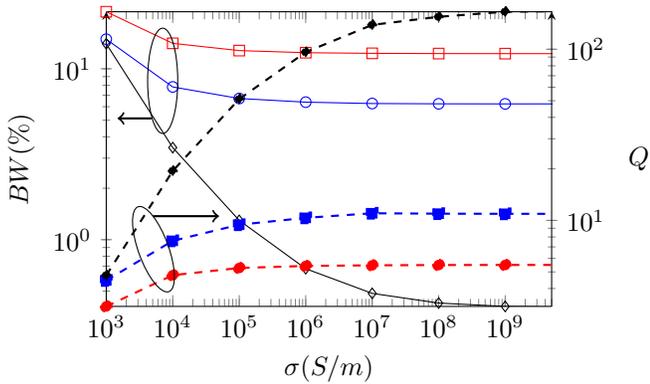
% % % % % % % % % % % % % % % % % % % % % % % % % % % % % % % % % % % % % % % % % % % % % % % % % % % % 
% % % % % % % % % % % % % % % % % % % % % % % % % % % % % % % % % % % % % % % % % % % % % % % % % % % % 

\section{Fabrication of the prototypes}
\label{Sec_Fabrication}

\begin{figure}
\includegraphics[width=0.95\linewidth]{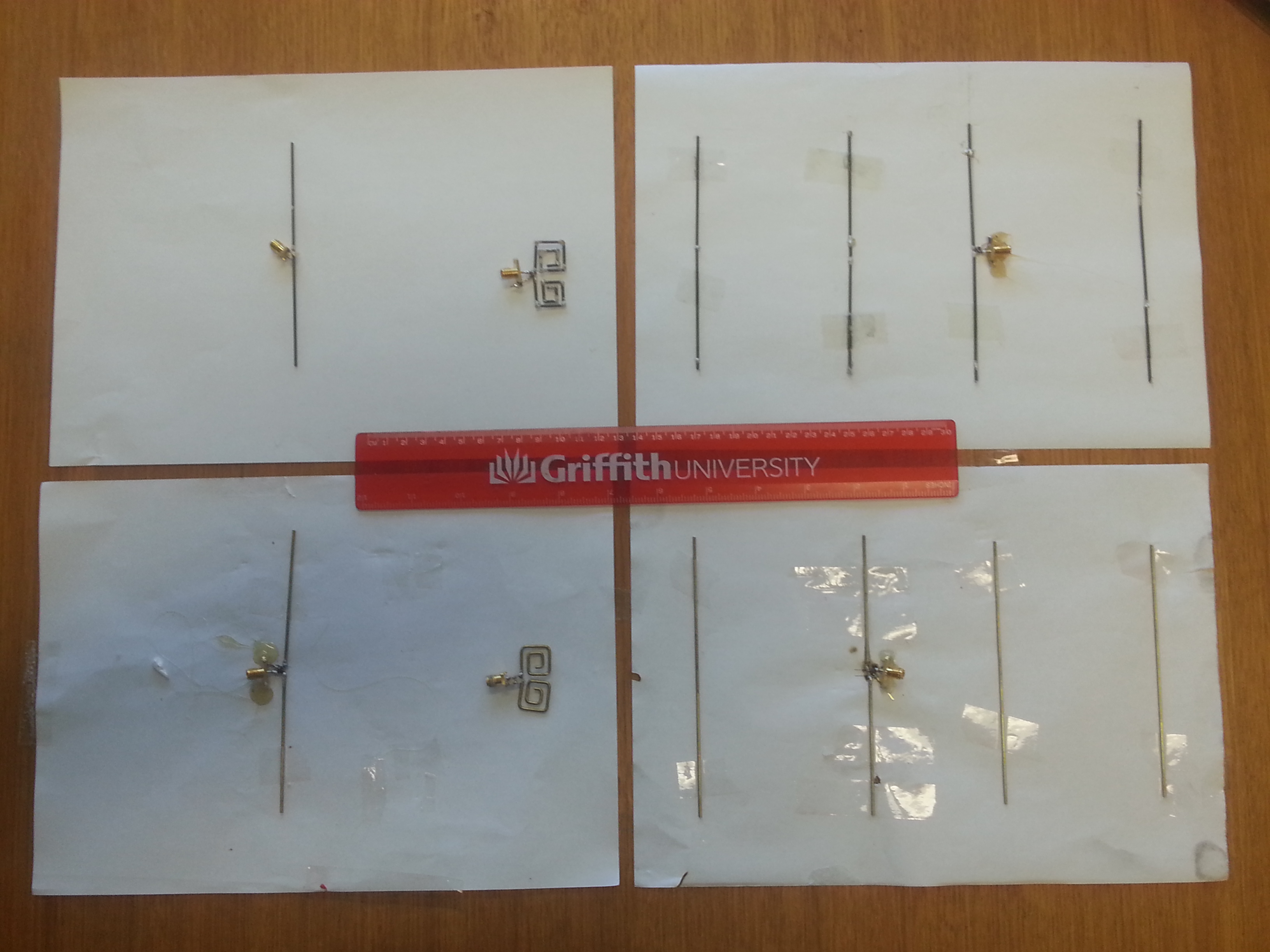}
\caption{Picture of fabricated antennas with graphite (top) and brass (bottom), antennas from left to right in each row: dipole, meander line, Yagi-Uda}
\label{Fig_AntPrototype}
\end{figure}

Fabrication of the antennas with conductivities much less than good conductors (like copper) is a challenge.
One of the main difficulties is soldering such low conductive materials.
In this study, we used brass and graphite materials for fabrication of the prototypes.
Brass has a conductivity of \SI{2.56e7}{\siemens\per\meter} and shows good soldering capabilities.
Pure graphite is strongly anisotropic material with significantly different conductivities in different planes ($\sigma_\perp \approx 3.3\times 10^2$ and $\sigma_\parallel\approx 2-3\times 10^5$S/m) \cite{Graphite_hb}.
Commercial leads of mechanical pencils (with \SI{1.4}{\milli\meter} diameter, and \SI{6}{\centi\meter} long) were used as the graphite rods.

\begin{figure*}
	\includegraphics[width=0.87\linewidth]{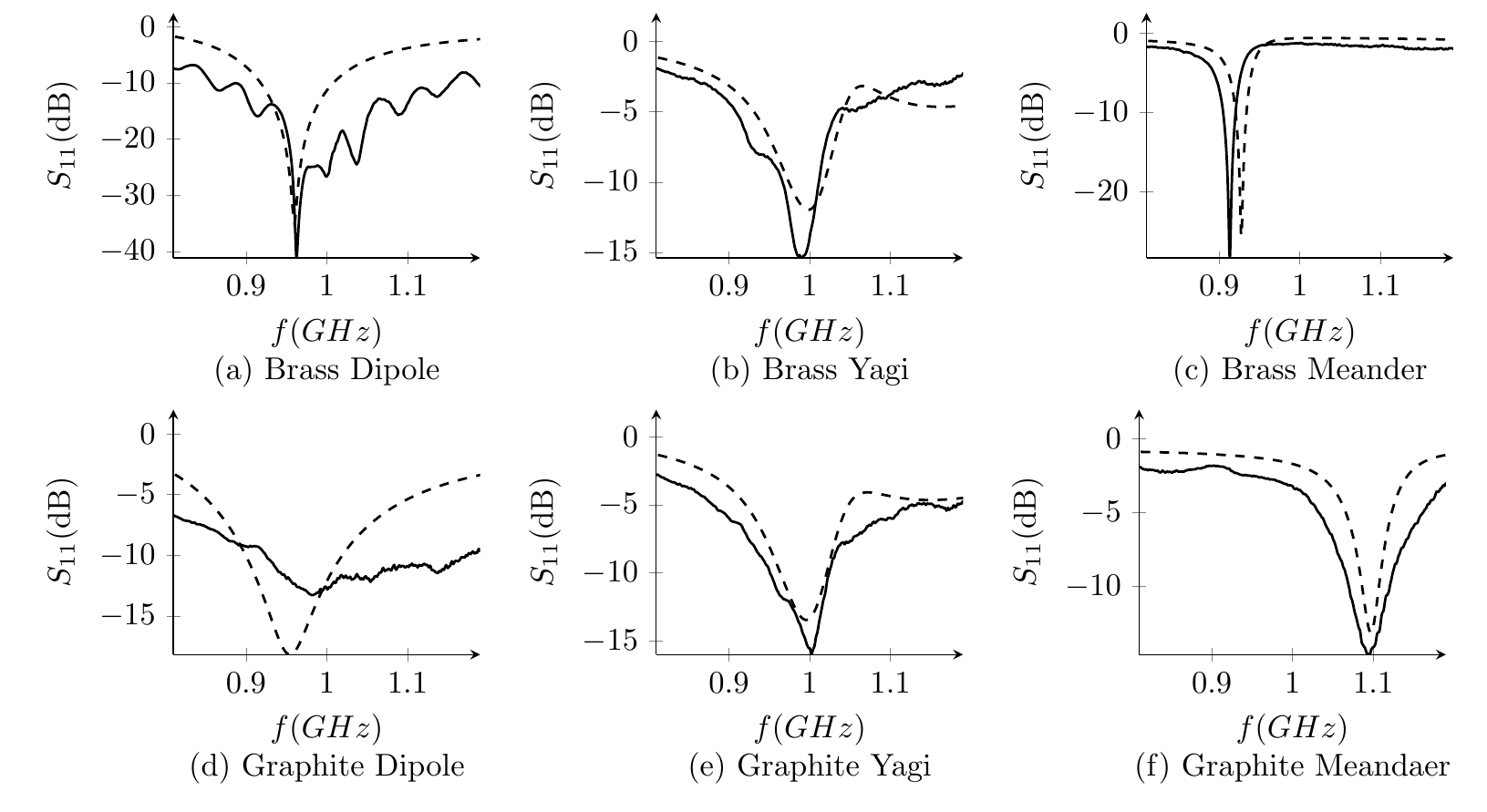}
	\caption{{Reflection coefficient of the fabricated antenna prototypes.
	Dashed and solid lines denote the simulations and measurement results, respectively.}}
	\label{Fig_Meas_s11_etarad}
\end{figure*}

We used conductive epoxy to make electrical connections  with graphite.
Sample antennas are depicted in  Fig.~\ref{Fig_AntPrototype}.
A chip is added as balun and transformer in all of the antennas to adjust the level of input impedance.
As the graphite rods are fragile, we fixed them with tape glue on a white paper placed on top of flat foam.

\section{Measurement results}
\label{Sec_MeasRes}

\begin{figure*}
\includegraphics[width=0.9\linewidth]{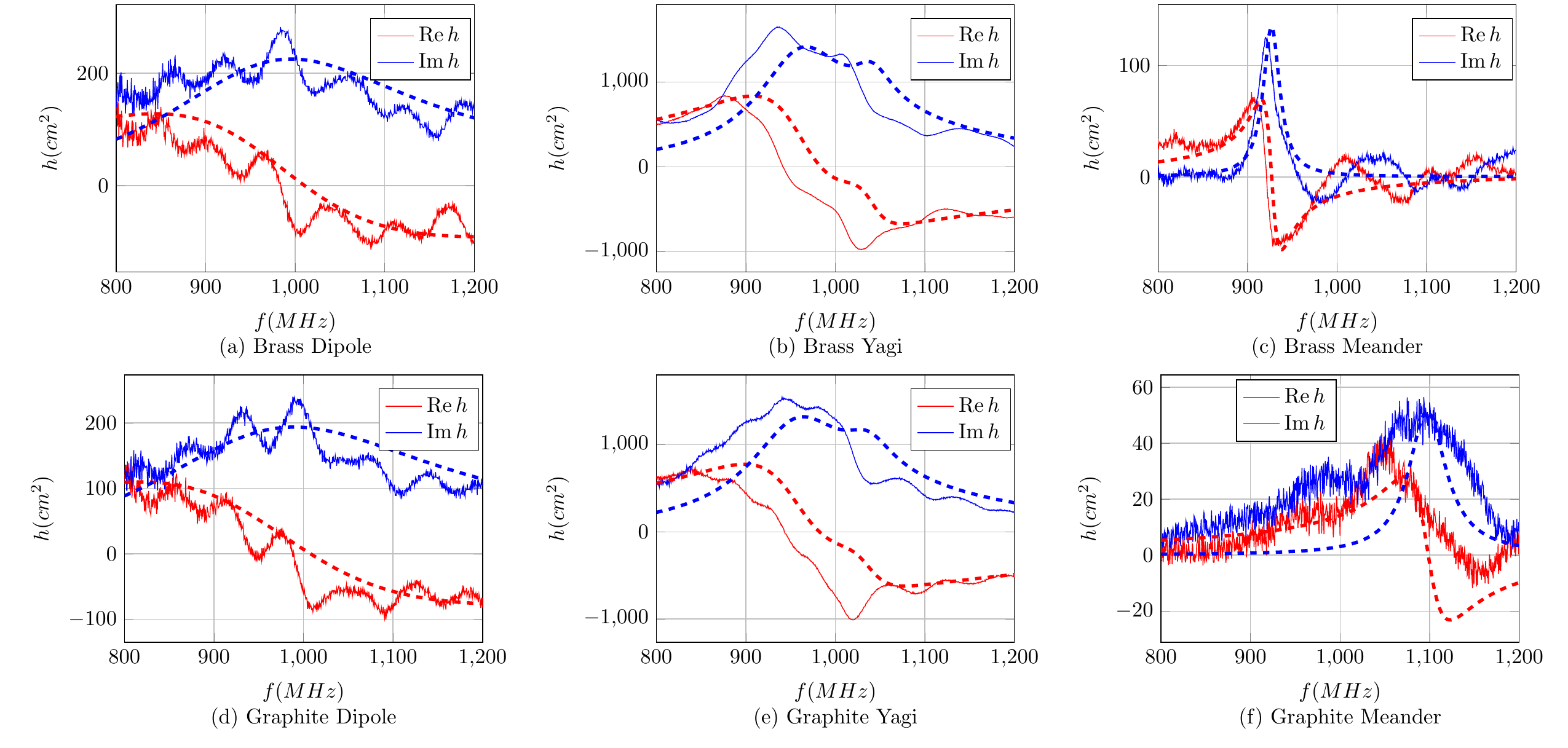}
\caption{Simulated and measured real and imaginary parts of the scattering coefficient $h$ (by optical theorem $\sigma_{ext}=\Im h$). 
Dashed and solid lines denote the simulations and measurement results, respectively.}
\label{Fig_Meas_sigma_ext}
\end{figure*}

	Each of the measured parameters is reported along with simulations for the sake of comparison.
	Brass antennas were modeled with conductivity of \SI{2.56e7}{\siemens\per\meter}.
	The graphite conductivity was measured at DC (explained in Appendix~\ref{App_MeasCond}) since we did not have access to RF material measurement setup.
	This might be the source of some difference between simulation and measurement results.

Illustrated in  Fig.~\ref{Fig_Meas_s11_etarad} are the simulated and measured scattering parameters $S_{11}$ for the fabricated antennas.
It is seen that Yagi-Uda resonates at \SI{1}{\giga\hertz} while brass and graphite meander lines have $5\%$ and $10\%$ shift in resonant frequency from the desired \SI{1}{\giga\hertz} resonance.
The reason for the frequency shift is that the SMA connector and cable have non-negligible effects on the meander line which should be included in the modeling.
The balun-transformer used to adjust low radiation resistance of the meander lines was included in the simulations.
The graphite was so fragile that it was not possible to use high precision machinery to shape the meander with exact dimensions which left us with manual alignment of the graphite parts. 
Therefore, the geometry of the meander was subject to human error, which we tried to measure  and include in the modeling to achieve a good agreement between simulation and measurement.

Measurement of the extinction cross section $\sigma_{ext}$ was performed using the optical theorem \cite[Ch.~10]{Ishimaru_1991_electromagnetic}.
This relates the extinction cross section of an arbitrary scatterer to the imaginary part of the scattering amplitude of the object in the forward direction $f(\hat{i},\hat{i})$ as:
\begin{align}
\sigma_{ext} = \frac{4\pi}{k}\Im f(\hat{i},\hat{i}) \cdot \hat{e}_i. 
\end{align}
where $\Im$ denotes the imaginary part and  $\hat{e}_i$ is a unit vector in the direction of polarization of incident wave.
Following the procedure outlined in \cite{Gustafsson_2012_TAP_FWScat}, we measured the signal at the receiver end once without the presence of antenna under test(AUT) $E_{r,0}$.
Antenna then was placed in the middle of the transmitter and receiver (distance $d$ from each), and the received signal recorded again as $E_{r,s}$.
Therefore, $\sigma_{ext}$ is found as
\begin{align}
\sigma_{ext} = \Im h \quad \mathrm{where} \quad h=\frac{2\pi d} {k}\left[\frac{E_{r,s}}{E_{r,0}} -1\right]^\ast,
\label{Eq_Sigma_Ext}
\end{align}
where $\ast$ denotes the complex conjugate operator.
It should be noted that propagating wave was assumed as $\exp{[-i(\omega t -kr)]}$ in \cite{Gustafsson_2012_TAP_FWScat}, however, we assumed $\exp{[+i(\omega t -kr)]}$ in our experimental setup. 
That is, $h$ in \eqref{Eq_Sigma_Ext} is the complex conjugate of  relation (17) in \cite{Gustafsson_2012_TAP_FWScat} and avoids negative cross section values \cite{Andersen_PersonalComm}.

Real and imaginary part of $h$ are illustrated in  Fig.~\ref{Fig_Meas_sigma_ext}.
As the Yagi-Uda occupies a larger area, it has larger extinction cross section while dipole and meander line have lower cross sections.
Fluctuations in the $\Re h$ and $\Im h$ are not observed for the Yagi-Uda antenna, as the Yagi-Uda has aperture 5 to 10 times larger than dipole and meander line.
Simulated extinction cross sections are included in the Fig.~\ref{Fig_Meas_sigma_ext} for the comparison which show good agreement with measurement except for the graphite meander.
We believe that in addition to human error in the apparatus (e.g. align phase center of AUT in the middle of transmitter and receiver) the conductive epoxy junctions could be another source of error.
These junctions have much higher conductivities than graphite.
Their primary function is to allow the flow of the current from one part to another.
However, when illuminated by the incoming wave, the conductive glue junctions might also act as the point scatterers.
As the graphite meander has more than 16 of such junctions, their might have significant impacts on the cross section measurements, although one cannot avoid or compensate their effect.

\begin{table}
\centering
\caption{Summary of the measurement results}
\label{Table_meas_summary}

% Please add the following required packages to your document preamble:
% \usepackage{booktabs}
% \usepackage{multirow}
% \usepackage{graphicx}
% \usepackage[table,xcdraw]{xcolor}
% If you use beamer only pass "xcolor=table" option, i.e. \documentclass[xcolor=table]{beamer}

\newcommand{\ra}[1]{\renewcommand{\arraystretch}{#1}}
\ra{1.3}
\begin{tabular}{@{}cccccc}
\toprule
              & \multicolumn{2}{c}{\textbf{Simulation}}       &\phantom{abc}& \multicolumn{2}{c}{\textbf{Measurement}}\\
              \cmidrule{2-3} \cmidrule{5-6}
              & $\eta_r$ & $\sigma_{ext} (cm^2)$ &\phantom{abc}&$\eta_r$ & $\sigma_{ext} (cm^2)$ \\
\midrule
\textbf{Brass}\\
Dipole       &	0.9859  &	251.8	&\phantom{abc}& 0.9281   & 282.3                 \\
Yagi         &	0.9893	&	1204	&\phantom{abc}& 0.7895   & 1647                  \\
Meander line &	0.6937	&	146.58	&\phantom{abc}& 0.6864   & 125.1                 \\
\textbf{Graphite}         \\
Dipole       &  0.741   &	213.5	&\phantom{abc}& 0.842    & 239     			 \\
Yagi         &	0.8194	&	1169	&\phantom{abc}& 0.7084   & 1524                  \\
Meander line &	0.2527	&	49.35	&\phantom{abc}& 0.3382   & 100.6                 \\
         \bottomrule
\end{tabular}
\end{table}

We calculated the radiation efficiency of the antennas with the improved Wheeler Cap (IWC) method introduced by Johnston and McRory \cite{Johnston_1998_APM}.
The original Wheeler cap method \cite{Wheeler_1959_ProcIRE} assumes an \emph{R-L-C} equivalent circuit to estimate $\eta_r$ for resonant antennas.
The new method offers greater accuracy over the original Wheeler cap method \cite{Wheeler_1959_ProcIRE} and solves some issues of the Wheeler caps \cite{Mendes_2007_EuCAP,McKinzie_1997_APS}.
Calculated values for the $\eta_r$ are reported in Table.~\ref{Table_meas_summary}.

Measurement results are summarized in Table.\ref{Table_meas_summary}, and also compared with the simulation results for brass and graphite.
In each class of material, the dipole and meander line have the highest and lowest radiation efficiencies which is consistent with Fig.~\ref{Fig_eta_r_sigma}.
As the graphite is less conductive than brass, graphite antennas are less efficient with lower cross sections, and absorb less power.
It is interesting to note that the cross section results closely follow $\eta_r$ variations from the brass to graphite.

\section{Conclusion}

Three common wire antenna structures (half-wave dipole, Yagi-Uda, and meander line antennas) were studied to investigate the impact of finite conductivity on the transmitting and receiving performance. 
The bandwidth, $Q$ factor, directivity, gain, radiation efficiency, absorption efficiency, generalized absorption efficiency, and absorbed power to volume ratio  of the antennas were compared for materials with conductivities ranging from $10^3$S/m to PEC. 
It was illustrated that the absorption efficiency of some antennas does not change with conductivity, however, the absorption cross section of the antennas follows the behavior of the radiation efficiency.
The variation in total extinct power from the lossy antenna with radiation efficiency is evident.

Brass and graphite were used to make dipole, Yagi-Uda, and meander line antennas.
In addition to the return loss of the antenna, we also measured the radiation efficiency and  extinction cross section of the antennas by improved Wheeler cap and optical theorem, respectively.
Measured radiation efficiency and $Q$ factor are also reported in this paper.
Radiation efficiency  directly influences  measured absorbed power and $Q$ factor.

\appendices
\section{Accuracy of MoM}
\label{App_MoMAcuracy}
The penetration depth $\delta$ (caused by skin effect) is comparable with the wire radius $r$ at the lowest conductivity $10^3$S/m. 
The accuracy of the simulation models is satisfied  when $\sigma\gg \omega \epsilon_0$, and $r>\delta$ which are imposed in the software manual.
To ensure about the accuracy of the simulations, we simulated a low conductive dipole antenna  ($\sigma=10^3$S/m) with FEM and FDTD solvers from a commercial vendor~\cite{CST_MW}.
A comparison of the resonant frequency, radiation efficiency, directivity and gain  is illustrated in Table.~\ref{Table_Feko_Validation}.

\begin{table}
\centering
\caption{{Validation of MoM results through comparison with FEM and FDTD solvers (test case: dipole antenna with $\sigma=10^3$S/m)}}
\label{Table_Feko_Validation}
\newcommand{\ra}[1]{\renewcommand{\arraystretch}{#1}}
\ra{1.3}
\begin{tabular}{rrrr}
\toprule
			&	MoM			&	FEM		&	FDTD	\\
			\midrule
$f_r$(MHz)	& 	$936$		&	$924$	&	$927$	\\
$\eta_r$	&	$0.541$		&	$0.551$	&	$0.548$\\
$D$(dBi)	&	$2.114$		&	$2.147$	&	$2.159$\\	
$G$(dBi)	&	$-0.549$	&	$-0.444$&	$-0.448$\\
\bottomrule
\end{tabular}

\end{table}

\section{Measurement of Conductivity}
\label{App_MeasCond}
\begin{figure}
\includegraphics[width=0.9\linewidth]{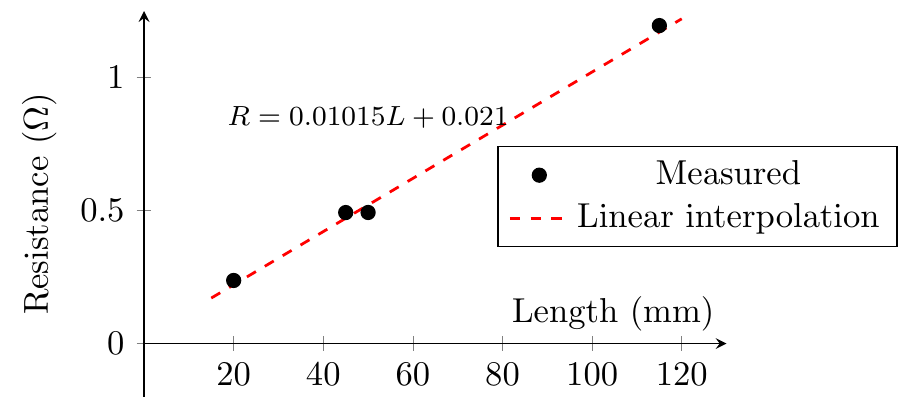}
\label{Fig_Conductivity_vs_L}
\caption{{Measurement of the resistance of several graphite rods (diameter 1.4mm) with different length (20,45,50,115mm).}}
\end{figure}

To measure the conductivity of the graphite rods, we measured the $V-I$ curve of graphite rods with the lengths of (20, 45, 50, 115mm) using an Agilent B1505A parameter analyzer. 
	All rods showed a linear $V-I$ curve. 
	The resistance of each rod was calculated from the slope of the $V-I$ curve. 
	Fig.~\ref{Fig_Conductivity_vs_L} shows the variation of the resistance versus the length. 
	A linear interpolation analysis provides $R=0.01015L+0.021$ as the best fit to the data (Pearson correlation coefficient $R^2=0.996$).
	Using the slope of the $R-L$ curve and cross sectional area of the rods, conductivity of the graphite was calculated as $\sigma=6.399 \times 10^4$(S/m)  which is 0.11\% of the conductivity of copper.
	
\section*{Acknowledgment}
Authors are grateful to Prof. J. B. Andersen for fruitful discussions on antenna scattering measurements.
Authors thank Mohammad Vatankhah, Dorian Oddo and Hanan Hamid from  Griffith School of Engineering for the help in the fabrication of the prototypes.
Assistance from Sasan Ahdirezaeieh of University of Queensland for measurement of extinction cross section of meander line graphite antenna is appreciated.
Graphite resistance measurements were performed by Dr P. Tanner at the Griffith University node of the Australian National Fabrication Facility.

\bibliographystyle{IEEEtran}
%\bibliography{IEEEabrv,library,Misc_Bib}
% Generated by IEEEtran.bst, version: 1.13 (2008/09/30)

\begin{IEEEbiography}[{\includegraphics[width=1in,height=1.25in,clip,keepaspectratio]{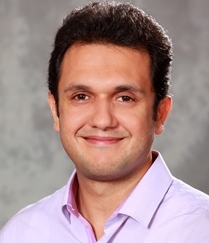}}]{Morteza Shahpari}
(S'08-M'15) received his bachelor and master degrees in telecommunications engineering from Iran University of Science and Technology (IUST), Tehran, Iran in 2005 and 2008, respectively.
He joined Griffith University, QLD, Australia in 2012 as a PhD candidate. He is currently a research fellow at Griffith School of Engineering.

His research interests include fundamental limitations of antennas, equivalent circuits for receiving antennas, antenna scattering.
\end{IEEEbiography}

\begin{IEEEbiography}[{\includegraphics[width=1in,height=1.25in,clip,keepaspectratio]{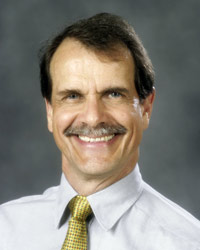}}]{David V. Thiel}
(SM'88) received a degree in physics and applied mathematics from the University of Adelaide, Adelaide, SA, Australia, and the master's and Ph.D. degrees from James Cook University, Townsville, QLD. He is currently the Deputy Head of School (Research) with Griffith University, Brisbane, QLD, Australia. He is a Fellow of the Institution of Engineers, Australia, and a Chartered Professional Engineer in Australia.

His research interests include electromagnetic geophysics, sensor development, electronics systems design and manufacture, antenna development for wireless sensor networks, environmental sustainability in electronics manufacturing, sports engineering, and mining engineering. He has authored the book Research Methods for Engineers and co-authored a book on Switched Parasitic Antennas for Cellular Communications.

Prof. Thiel has authored six book chapters, over 120 journal papers, and has co-authored more than nine patent applications. He was a co-inventor of the new RoHS and WEEE compliant electronics manufacturing technology called circuits in plastic. He is a regular reviewer for many engineering journals, and an Associate Editor of the International Journal of RF and Microwave Computer Aided Engineering and the IEEE Antennas and Propagation Magazine.	
\end{IEEEbiography}

\end{document}

\title{Supplementary information for  ``Impact of reduced conductivity on the performance of wire antennas''}

\author{Morteza~Shahpari,~\IEEEmembership{Student~Member,~IEEE,} David~V.~Thiel,~\IEEEmembership{Senior~Member,~IEEE,} \thanks{M. Shahpari and D. Thiel are with Centre for Wireless Monitoring and Applications, School of Engineering, Griffith University, Nathan, Queensland 4111, Australia} \thanks{This work is partly funded by the grant number  DP130102098 from Australian Research Council.}\thanks{Manuscript received xx, xx, xxxx; revised xx, xx, xxxx.}}

\markboth{IEEE Transactions on Antennas and Propagation,~Vol.~X, No.~X, January~20XX}{}

\maketitle

\section{Equivalent Circuit of Yagi-Uda antenna}
\label{Sec_SI_EqYagi}
\begin{figure}
\subfigure[Transmitting case]{
\includegraphics[width=0.5\textwidth]{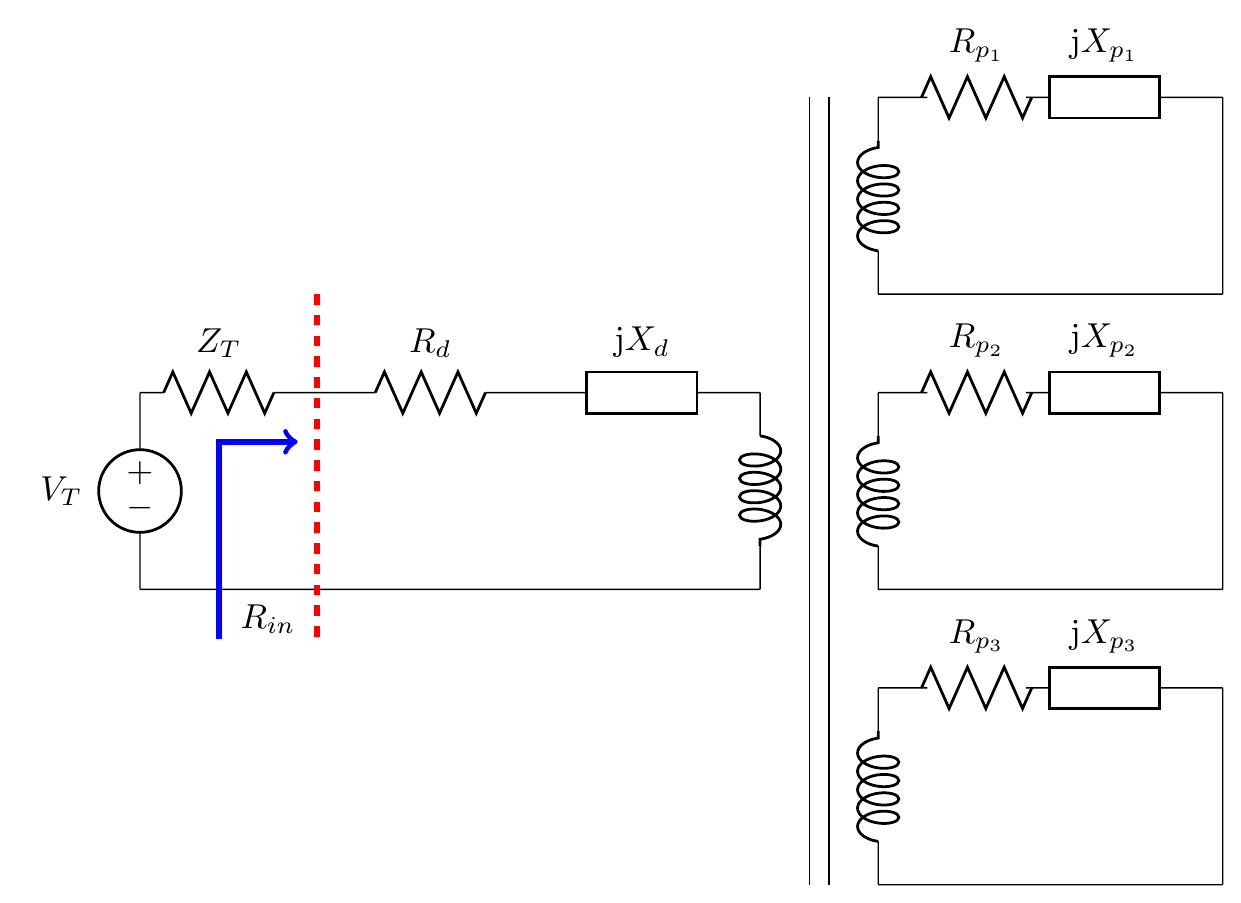}
}
\subfigure[Receiving case]{
\includegraphics[width=0.5\textwidth]{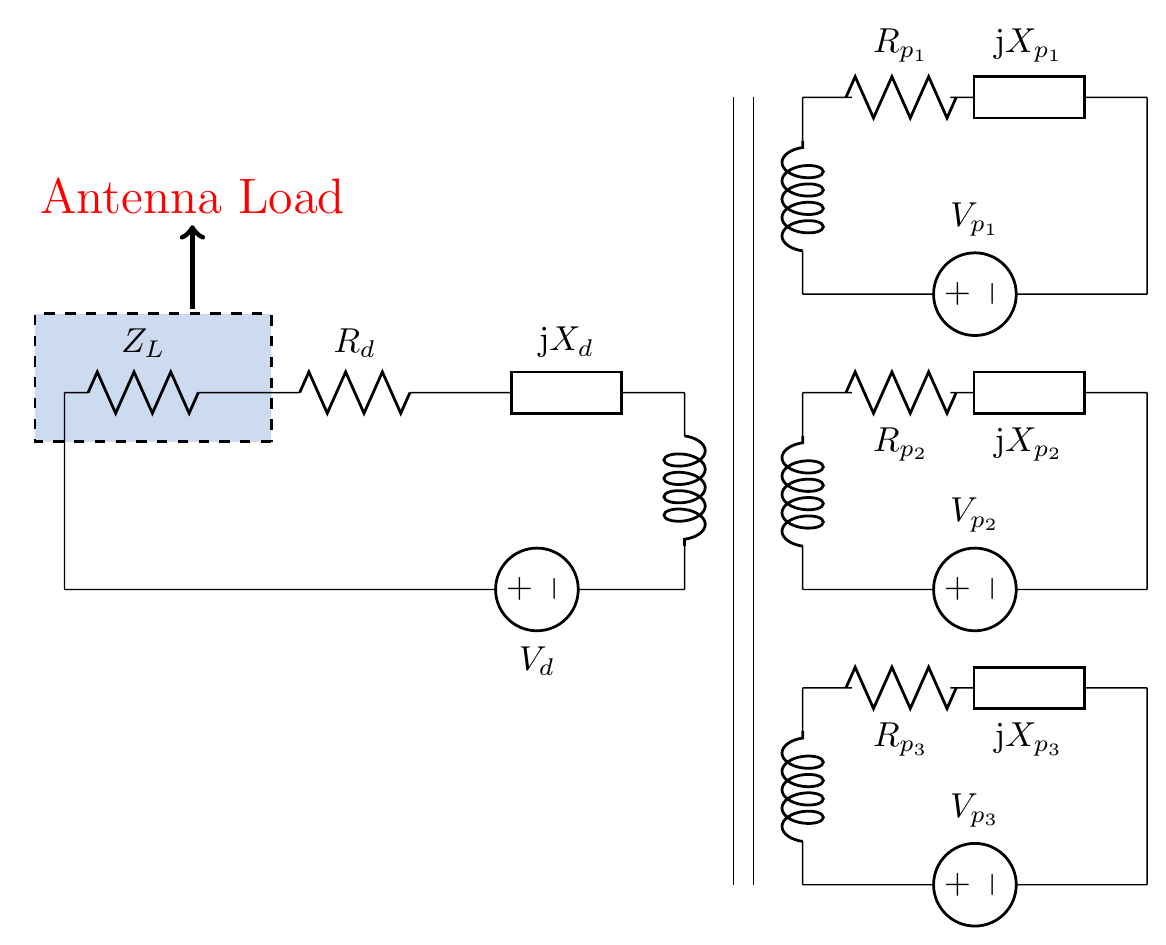}
}
\caption{Equivalent circuit of the Yagi-Uda antenna with 1 reflector and 2 directors}
\label{SI_Fig_Yagi_EqCir}
\end{figure}

Andersen and Vaughan \citep{Andersen_2003_APM} considered the impact of the parasitic elements in the antenna equivalent circuit.
They generalized Thevenin-Norton equivalent circuits to a multi port network for Yagi-Uda antennas with one port allocated to each parasitic element.
Fig.S\ref{SI_Fig_Yagi_EqCir} shows a sample equivalent circuit of the Yagi-Uda antenna with three parasitic elements.
Ports at the terminals of the parasitic elements are considered as short circuits while a voltage source is assumed for the receiving antenna.
An ideal transformer is considered to convert impedance of the parasitic elements to the driver side.

\section{Examination of Eq.\ref{Eq_S_Box}}
\label{Sec_SI_Graphs}
\begin{figure*}

%%\subfigure[Dipole]{
%%\begin{tikzpicture}
%%\pgfplotsset{every axis legend/.append style={
%%at={(0.95,.1)},
%%anchor=south east}}
%%\begin{semilogxaxis}[width = \FigWidth, height = \FigHeight,xlabel=$\sigma (S/m)$, ylabel=$G(dBi)$]
%%\addplot [only marks,mark = o] table[x = sigma, y expr = \thisrow{Dipole} ] {\DataRadEffdB};
%%\addplot [dashed, mark=square] table[x=sigma,y=Dipole] {\DataGdBi};
%%\legend{$\eta_r   D_{PEC}$, simulated $G$};
%%\end{semilogxaxis}
%%\end{tikzpicture}
%%}
%%\subfigure[Yagi]{
%%\begin{tikzpicture}
%%\pgfplotsset{every axis legend/.append style={
%%at={(0.95,.1)},
%%anchor=south east}}
%%\begin{semilogxaxis}[width = \FigWidth, height = \FigHeight, ,xlabel=$\sigma (S/m)$, ylabel=$G(dBi)$]
%%\addplot [only marks,mark = o] table[x = sigma, y expr = \thisrow{Yagi} ] {\DataRadEffdB};
%%\addplot [dashed, mark=square] table[x=sigma,y=Yagi] {\DataGdBi};
%%\legend{$\eta_r   D_{PEC}$, simulated $G$};
%%\end{semilogxaxis}
%%\end{tikzpicture}
%%}
%%\subfigure[Meander]{
%%\begin{tikzpicture}
%%\pgfplotsset{every axis legend/.append style={
%%at={(0.95,.1)},
%%anchor=south east}}
%%\begin{semilogxaxis}[width = \FigWidth, height = \FigHeight, ,xlabel=$\sigma (S/m)$, ylabel=$G(dBi)$]
%%\addplot [only marks,mark = o] table[x = sigma, y expr = \thisrow{Meander}  ] {\DataRadEffdB};
%%\addplot [dashed, mark=square] table[x=sigma,y=Meander] {\DataGdBi};
%%\legend{$\eta_r   D_{PEC}$, simulated $G$};
%%\end{semilogxaxis}
%%\end{tikzpicture}
%%}
\includegraphics[height=\FigHeight]{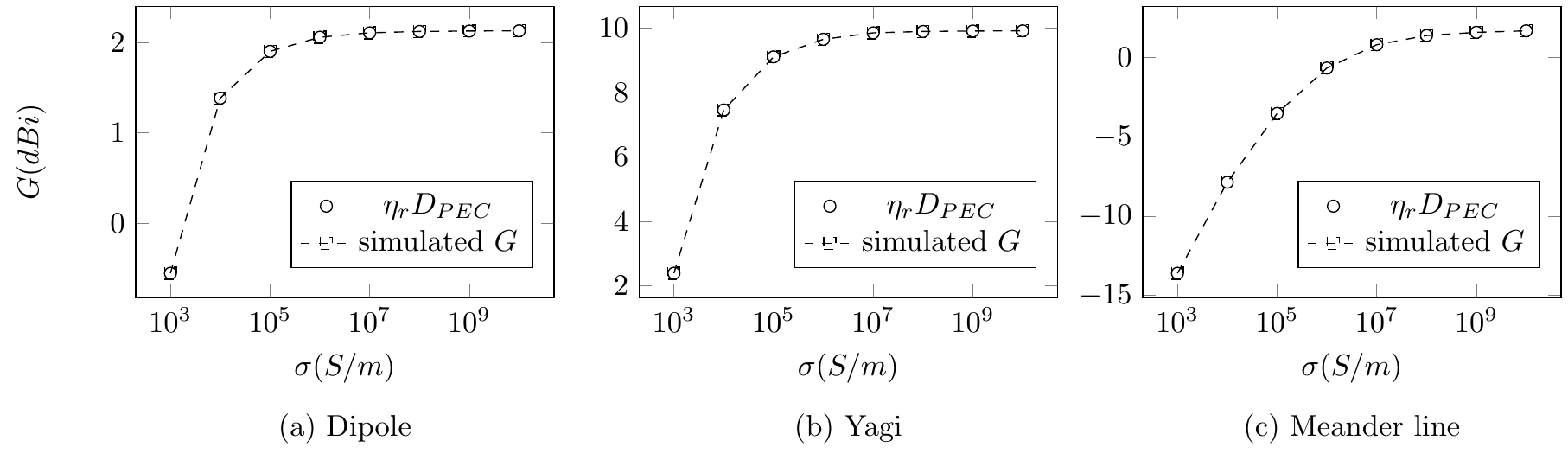}
\caption{Investigation of $G=\eta_r D$ relation for antennas with different conductivities}
\label{SI_Fig_G_eta_r}
\end{figure*}

\begin{figure*}

%%%\begin{tikzpicture}
%%%\pgfplotsset{every axis legend/.append style={
%%%at={(0.95,.1)},
%%%anchor=south east}}
%%%\begin{semilogxaxis}[width = \FigWidth, height = \FigHeight,,xlabel=$\sigma (S/m)$, ylabel=$Q$]
%%%\addplot [mark = o,only marks] table[x = sigma, y expr = \thisrow{Dipole} * 5.4912 ] {\DataRadEff};
%%%\addplot [dashed, mark=square] table[x=sigma,y=Dipole] {\DataQ};
%%%\legend{$\eta_r   Q_{PEC}$, simulated $Q$};
%%%\end{semilogxaxis}
%%%\end{tikzpicture}
%%%\begin{tikzpicture}
%%%\pgfplotsset{every axis legend/.append style={
%%%at={(0.95,.1)},
%%%anchor=south east}}
%%%\begin{semilogxaxis}[width = \FigWidth, height = \FigHeight, ,xlabel=$\sigma (S/m)$, ylabel=$Q$]
%%%\addplot [mark = o,only marks] table[x = sigma, y expr = \thisrow{Yagi} * 10.9251 ] {\DataRadEff};
%%%\addplot [dashed, mark=square] table[x=sigma,y=Yagi] {\DataQ};
%%%\legend{$\eta_r   Q_{PEC}$, simulated $Q$};
%%%\end{semilogxaxis}
%%%
%%%\end{tikzpicture}
%%%\begin{tikzpicture}
%%%\pgfplotsset{every axis legend/.append style={
%%%at={(0.95,.1)},
%%%anchor=south east}}
%%%\begin{semilogxaxis}[width = \FigWidth, height = \FigHeight, ,xlabel=$\sigma (S/m)$, ylabel=$Q$]
%%%\addplot [mark = o,only marks] table[x = sigma, y expr = \thisrow{Meander} * 167.0969 ] {\DataRadEff};
%%%\addplot [dashed, mark=square] table[x=sigma,y=Meander] {\DataQ};
%%%\legend{$\eta_r   Q_{PEC}$, simulated $Q$};
%%%\end{semilogxaxis}
%%%\end{tikzpicture}
\includegraphics[height=\FigHeight]{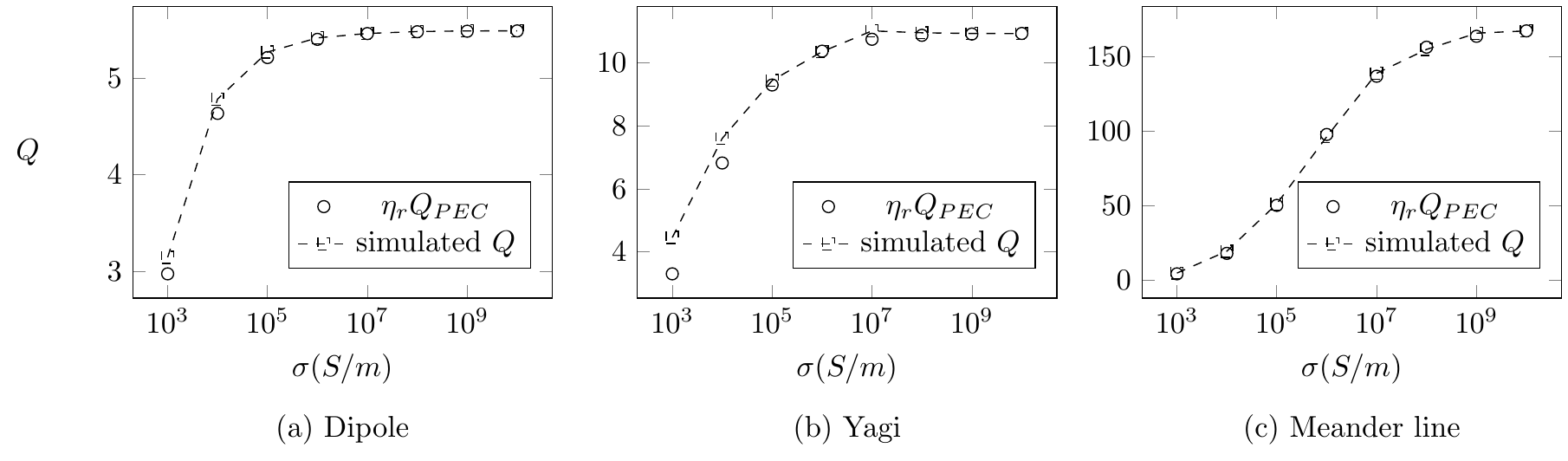}
\caption{Variation of the $Q$ factor with and radiation efficiency $\eta_r$}
\label{SI_Fig_Q_eta_r}
\end{figure*}

It is the aim of this supplementary section to provide additional data  for examination of the validity of relations in the following form:
\begin{align}
\Box_{lossy} = \eta_r \Box_{PEC}.
\label{Eq_S_Box}
\end{align}
Fig.S\ref{SI_Fig_G_eta_r}-S\ref{SI_Fig_Q_eta_r}  demonstrate if gain $G$, $Q$ factor, absorption, scattering, extinction cross section, repectively. $\sigma_a$, $\sigma_s$, and $\sigma_{ext}$ follow the behaviour of the radiation effiency or not.
For example, Fig.S\ref{SI_Fig_Q_eta_r} investigates the relation of $Q$ factor of a lossy and lossless antenna as $Q_{lossy} = \eta_r   Q_{PEC}$.

\bibliographystyle{IEEEtranN}
%\bibliography{IEEEabrv,ConductivityRef}

% Generated by IEEEtranN.bst, version: 1.13 (2008/09/30)

% % % % % % % % % % % % % % % % % % % % % % % %

% % % % % % % % % % % % % % % % % % % % % % % % % %
\begin{figure*}

\includegraphics[height=\FigHeight]{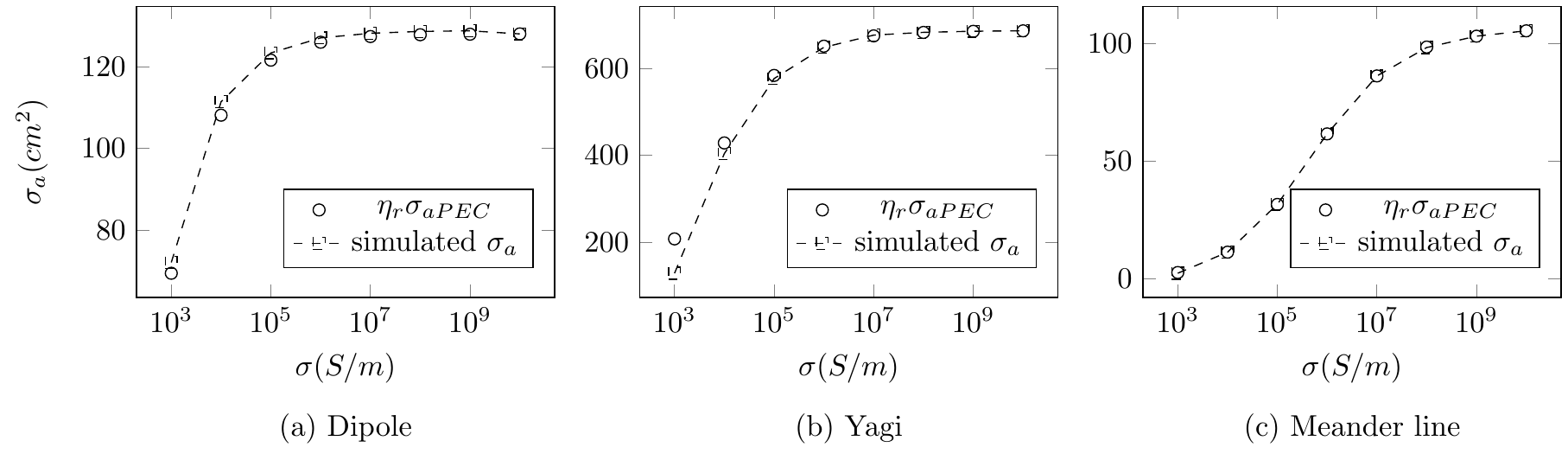}
\caption{Variation of the absorption cross section $\sigma_a$ and radiation efficiency $\eta_r$}
\label{SI_Fig_SigmaAB_eta_r}
\end{figure*}

% % % % % % % % % % % % % % % % % % % % % % %

\begin{figure*}

\includegraphics[height=\FigHeight]{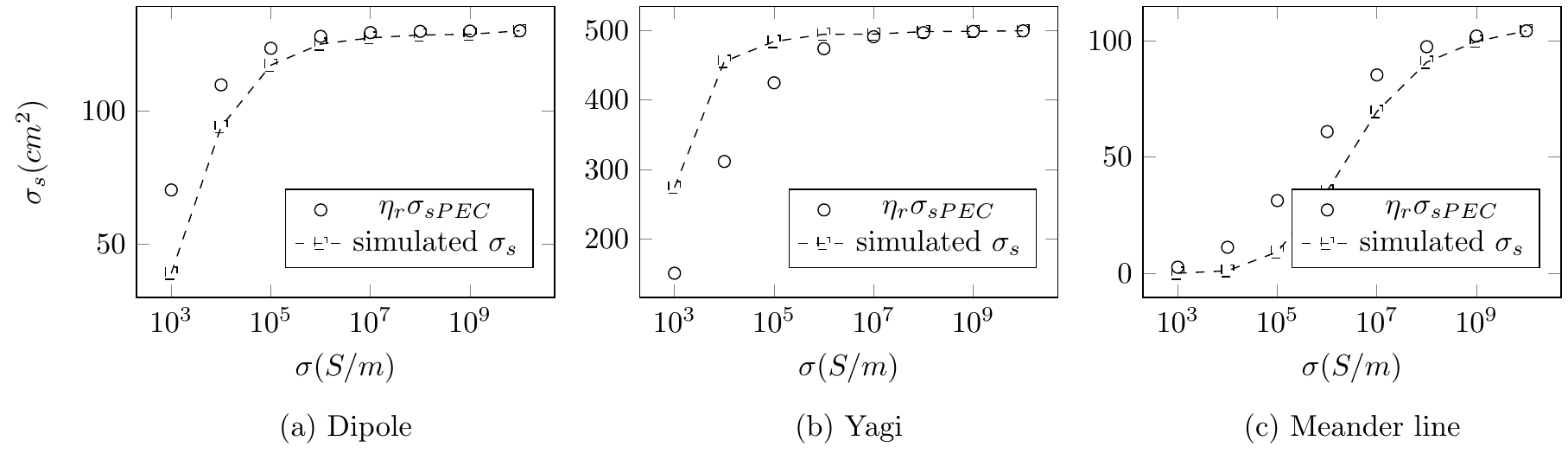}
\caption{Variation of the scattering cross section $\sigma_s$ and radiation efficiency $\eta_r$}
\label{SI_Fig_SigmaSC_eta_r}
\end{figure*}

% % % % % % % % % % % % % % % % % % % % % % %
\begin{figure*}
\includegraphics[height=\FigHeight]{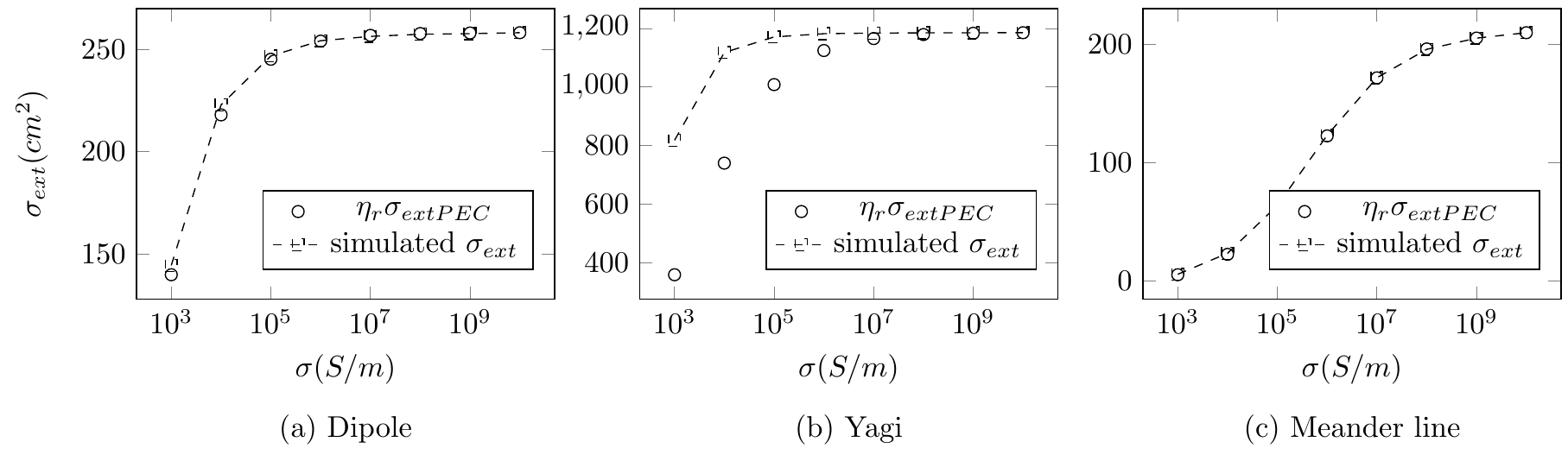}

\caption{Variation of extinction cross section $\sigma_{ext}$ and radiation efficiency $\eta_r$}
\label{SI_Fig_SigmaEXT_eta_r}
\end{figure*}

\begin{figure*}
\textcolor{white}{\rule{0.8\textwidth}{0.8\textwidth}}
\end{figure*}